\documentclass[11pt]{article}
\usepackage{graphicx}
\begin{document}
\title{\textbf{\textsf{Constraining the coupling constant between dark energy and dark matter}}}
\author{ Mubasher Jamil\footnote{mjamil@camp.edu.pk}\ \ and Muneer Ahmad Rashid\footnote{muneerrshd@yahoo.com}
\\ \\
%EndAName
\textit{\small Center for Advanced Mathematics and Physics}\\
\textit{\small National University of Sciences and Technology}\\
\textit{\small Peshawar Road, Rawalpindi, 46000, Pakistan} \\
%EndAName
} \maketitle \large
\begin{abstract}
\textsf{We have investigated constraints on the coupling between
dark matter and the interacting Chaplygin gas. Our results indicate
that the coupling constant $c$ between these two entities can take
arbitrary values, which can be either positive or negative, thus
giving arbitrary freedom to the inter-conversion between Chaplygin
gas and dark matter. Thus, our results indicate that the restriction
$0<c<1$ on the coupling constant occurs as a very special case. Our
analysis also supports the existence of phantom energy under certain
conditions on the coupling constant.}
\end{abstract}


\large
\section{Introduction}

It is well known that the expansion of the universe is accelerated.
It has been confirmed by numerous observations taken by various
scientific groups across the globe using WMAP \cite{sper}, distant
supernova type 1a data \cite{perl,riess}, large-scale structure and
galaxy distribution \cite{eisen} and the gravitational lensing
phenomenon of high-redshift galaxies \cite{Biesiada}. These
observations clearly suggest that the universe is spatially flat and
is dominated by some sort of vaccum energy having negative pressure
commonly called `dark energy'. This energy has been interpreted in
various forms like the cosmological constant \cite{wein,carro},
quintessence models based on the ideas of a spatially homogeneous
and time-dependent scalar field \cite{peebl}, phantom energy
\cite{cald,noji,noji1,jamil15,jamil16}, quintom models
\cite{feng,feng1,wu1}, k-essence \cite{Chimento}, holographic dark
energy \cite{seta} and the Chaplygin gas (CG) \cite{Bento1,zhang11}
(see also \cite{cope} for a recent review on dark energy).

The CG is represented by an equation of state (EoS) of the form
$p=-A/\rho ,$ where $A$ is a constant parameter \cite{Kamen}. The CG
gives rise to a simple cosmological model that interpolates between
the earlier matter-(or dust-) dominated to the later dark energy
dominated phase of the universe. Due to its effectiveness in
explaining the evolution of the universe, several generalizations of
CG have been proposed in the literature \cite{bena,debna,dev,sen}.
The observational evidence in support of cosmological models based
on the CG-EoS is also very encouraging \cite{Nesseris,bean}. The CG
also possesses the property of giving accelerated expansion even if
it gets coupled with other scalar fields like quintessence\ or
dissipative\ matter fields \cite{Chimento1}. It also yields
traversable wormhole solutions to the Einstein field equations if
the pressure and density of CG violates the null energy condition
\cite{lobo}. Besides its various useful implications in cosmology,
CG has the drawback of producing oscillations or exponential blow up
of dark matter power spectrum which is inconsistent with
observations \cite {sand}. Similar results are obtained in later
generalizations of CG \cite{Carturan}. But later it was proved that
such oscillations can be avoided and structure formation can proceed
(which is strongly supported by dark matter) if the phantom-like
dark energy is excluded, thereby proceeding with only dark matter
and dark energy \cite{Bento}. It was further suggested that CG
behaved like a passive background in the early evolution of the
universe and that only dark matter leads to nonlinear growth of
structures but later the evolution is dominated by CG \cite{Manera}.
The inhomogeneities generated by dark matter were stabilized by the
CG which is compatible with the observations \cite{Koivisto}. Later
studies on supernovae data put constraints on CG leading to
cosmological models based on CG to behave just like the cosmological
constant \cite{Avelino}.

Modern cosmology is plagued with numerous theoretical and
observational problems: among them is the \textit{cosmic-coincidence
problem} which can be stated thus \cite{jamil17}: why are the energy
densities of matter and dark energy almost of the same order at
present? In the standard cosmological model, the ratio of the energy
densities of matter and dark energy should fall rapidly as the
universe expands, but observationally the corresponding ratio turns
out to be almost constant or minutely fluctuating around unity, a
phenomenon commonly called the `soft coincidence'. It leads to the
possibility that energy might be exchanged to keep such a delicate
balance in the densities. This interaction is generally studied in
the models so-called `interacting dark energy'
\cite{jamil14,jamil15,jamil12,jamil13}. Unified models based on dark
matter and CG have been widely investigated (see \cite{Bento} and
references therein) but the fundamental question dealing with the
interaction between these two entities is not satisfactorily
answered and requires further investigation. A cosmological model
based on the interacting CG had been proposed \cite{zhang} to
investigate this interaction. This model yields the result that the
universe is to cross the phantom divide i.e. the transition from the
state $\omega>-1$ to $\omega <-1 $ or more simply $\omega =-1$,
which is not possible in the models based on pure CG. Furthermore,
this leads to the scaling solutions of the cosmological dynamical
system, which helps in explaining the coincidence problem
effectively. Moreover, it is used in cosmological models to
investigate dissipative effects of van der Waal's fluid and dark
energy \cite{kremer}. In fact it is later suggested that without
interaction with other species, any hydrodynamical or k-essence like
model in general relativity cannot cross $\omega=-1$ \cite{vik}. The
interacting CG also yields stable scaling solutions of
Friedmann-Lemaitre-Robertson-Walker (FLRW) equations at late times
of the universe. This model was later extended to the case of an
interacting generalized Chaplygin gas (GCG) \cite{wu}. There are
some proposals that this interaction can be observed if cubic
corrections are provided to the Hubble law, measured by distant
supernovae of type 1a \cite{szyd}. It is worthwhile to understand
the role of the coupling constant in the interacting models which we
have investigated using a modified Chaplygin gas (MCG).

The outline of this paper is as follows: In the next section, we
model our dynamical system on the pattern of \cite {wu} and
determine the critical points corresponding to that system. In the
third section, we perform stability analysis corresponding to each
critical point. In the fourth section, we perform analysis to
determine constraints on the coupling constant. Finally we present
conclusion of our paper.

\section{Modeling of dynamical system}

We start by assuming the background to be a spatially homogeneous
and isotropic FLRW spacetime,
\begin{equation}
ds^2=-dt^2+a^2(t)\left[ \frac{dr^2}{1-kr^2}+
r^2(d\theta^2+\sin^2\theta d\phi^2)\right],
\end{equation}
where $a(t)$ is the scale factor and the curvature parameter
$k=-1,0,+1$ describes spatially open, flat or closed spacetimes. It
is assumed that the spacetime is filled with the a two-component
fluid, namely dark matter and dark energy. The corresponding
energy-momentum tensors are specified by
\begin{equation}
T_{\mu\nu}^{(dm)}=\rho_{dm} u_\mu^\prime u_\nu^\prime, \ \
T_{\mu\nu}^{(mcg)}=(\rho_{mcg}+p_{mcg}) u_\mu u_\nu+p_{mcg}u_\mu
u_\nu.
\end{equation}
Here $u_\mu^\prime$ and $u_\mu$ is the comoving four-velocity of
dark matter and dark energy respectively. The notations $dm$ and
$mcg$ corresponds to dark matter the dark energy, respectively. In
our model, the dark energy is specified by the modified Chaplygin
gas EoS \cite{li}
\begin{equation}
p_{mcg}=A\rho _{mcg}-\frac{B}{\rho _{mcg}^{\alpha }},
\end{equation}%
where $A$ and $B$ are constant parameters and $0\leq\alpha \leq 1$.
The MCG reduces to GCG if $A=0$ and to the CG if furthermore $\alpha
=1.$ It reduces to the standard linear EoS for a perfect fluid if
$B=0.$ Recently it has been deduced, using latest supernova data,
that models with $\alpha>1$ are also possible \cite{berto}. For our
analysis the positivity of $\alpha$ is sufficient. In our further
discussion, the MCG and dark energy are used interchangeably.

The density evolution of MCG is given by
\begin{equation}
\rho _{mcg}=\left(\frac{B}{1+A}+\frac{C}{a^{3(1+A)(1+\alpha
)}}\right)^{\frac{1}{1+\alpha }},
\end{equation}
where $C$ is the constant of integration. Note that Eq. (4) holds
only when the interaction is absent. The equations of motion
corresponding to FLRW spacetime filled with a two-component fluid
are
\begin{eqnarray}
\dot{H}&=&-\frac{\kappa ^{2}}{6}(p_{mcg}+\rho _{mcg}+\rho _{dm}),\\
H^{2}&=&\frac{\kappa ^{2}}{3}(\rho _{mcg}+\rho _{dm}).
\end{eqnarray}
Here $\kappa ^{2}=8\pi G$ is the Einstein gravitational constant and
$H=H(t)$ is the Hubble parameter. We assume $k=0$, representing a
flat model of the universe. Furthermore, the energy conservation for
the two-component perfect fluid is obtained from
\begin{equation}
\nabla^\nu
T_{\mu\nu}=\nabla^\nu(T^{(dm)}_{\mu\nu}+T^{(mcg)}_{\mu\nu})=0.
\end{equation}
Here $\nabla_\nu$ refers to the covariant derivative with respect to
$x^\nu$ coordinate. Eq. (7) yields
\begin{equation}
\dot{\rho}_{mcg}+\dot{\rho}_{dm}+3H(p_{mcg}+\rho_{mcg}+\rho_{dm})=0.
\end{equation}
Due to interaction, the energy will not independently be conserved
for the interacting components, and therefore
\begin{equation}
\nabla^\nu T^{(mcg)}_{\mu\nu}=-Q_\mu,\ \ \nabla^\nu
T^{(dm)}_{\mu\nu}=Q_\mu.
\end{equation}
Here $Q_\mu$ is the interaction term that corresponds to energy
exchange between dark energy and dark matter. Solving Eqs. (9) using
(2), we obtain the so-called energy-balance equations corresponding
to MCG and dark matter as:
\begin{eqnarray}
\dot{\rho }_{mcg}+3H(p_{mcg}+\rho _{mcg})&=&-Q,\\
\dot{\rho }_{dm}+3H\rho _{dm}&=&Q .
\end{eqnarray}
The function $Q\equiv Q_t$, $\mu=t$ has dependencies on the energy
densities and the Hubble parameter, i.e. $Q(H\rho_{dm})$,
$Q(H\rho_{mcg})$ or $Q(H\rho_{dm},H\rho_{mcg})$ \cite{feng2}.
Because of the unknown nature of both dark energy and dark matter,
it is not possible to derive $Q$ from first principles. In order to
deduce a reasonable $Q$, we may expand like
$Q(H\rho_{dm},H\rho_{mcg})\simeq \alpha_{dm} H
\rho_{dm}+\alpha_{mcg}H\rho_{mcg}$. Since the coupling strength is
also not known, we may adopt just one parameter for our convenience;
hence we take $\alpha_{dm}=\alpha_{mcg}=c$ \cite{campo}. We here
choose the following coupling function $Q$ given by \cite{jamil15}:
\begin{equation}
Q =3Hc(\rho _{mcg}+\rho _{dm}).
\end{equation}
The choice of $Q$ is completely arbitrary but care must be taken
that it must satisfy the energy conservation (see Ref. \cite{quar}
for various other forms of $Q$). Here $c$ is the corresponding
coupling constant (also called the `transfer strength') for the
interaction. To study the dynamics of our system, we\ proceed by
setting
\begin{equation}
x=\ln a=-\ln (1+z),
\end{equation}
where $z$ is the redshift parameter. Moreover, the density and
pressure of MCG can be expressed by dimensionless parameters $u$ and
$v$ as follows:
\begin{equation}
u=\Omega _{mcg}=\frac{\rho _{mcg}}{\rho _{cr}}=\frac{\kappa ^{2}\rho
_{mcg}}{ 3H^{2}},\ \ v=\frac{\kappa ^{2}p_{mcg}}{3H^{2}}.
\end{equation}
The EoS parameter $\omega $ is conventionally defined as
\begin{equation}
\omega (x)\equiv \frac{p_{mcg}}{\rho _{mcg}},
\end{equation}
which becomes
\begin{equation}
\omega(x)=\frac{v}{u}.
\end{equation}
The density parameters of MCG and\ dark matter are related as
\begin{equation}
\Omega _{dm}=\frac{\kappa ^{2}\rho _{dm}}{3H^{2}}=1-\Omega
_{mcg}=1-u.
\end{equation}
Since $\Omega _{dm}\sim 0.3$ \cite{perl},\ it constrains $u\in
(0,1)$ for a flat universe. The following system of differential
equations governing the dynamics of our cosmological model is
determined by using the above equations:
\begin{eqnarray}
\frac{du}{dx}&=&-3c-3v+3uv,\\
\frac{dv}{dx}&=&-3\left[\frac{v}{u}+\left(A-\frac{v}{u}\right)(1+\alpha
)\right](u+v+c)+3v(1+v).
\end{eqnarray}
Note that for $A=0$, the above system reduces to the case for
interacting generalized Chaplygin gas \cite{wu}. By equating Eqs.
(18) and (19) to zero, we obtain the three critical points
$(u_{ic},v_{ic})$, with $i=1,2,3$, given by
\begin{eqnarray}
u_{1c}&=&1-c,\\
v_{1c}&=&-1,\\
u_{2c}&=&\frac{1}{2}-\frac{\sqrt{A+4c}}{2\sqrt{A}},\\
v_{2c}&=&\frac{1}{2}(A-\sqrt{A}\sqrt{A+4c}),\\
u_{3c}&=&\frac{1}{2}\left(1+\frac{\sqrt{A+4c}}{\sqrt{A}}\right),\\
v_{3c}&=&\frac{1}{2}(A+\sqrt{A}\sqrt{A+4c}).
\end{eqnarray}
For the critical points to be real valued, we require $A+4c\geq0$.
Notice that the first critical point is the same as discussed in
\cite{wu}. There it was proposed that the coupling constant $c\in
\lbrack 0,1].$ Since there is a transition from CG to dark matter
(i.e. $c\rightarrow 1$) as the universe evolves, it implies that the
future universe will contain only dark matter and might have no
trace of CG. Our analysis in the next two sections suggests that $c$
cannot necessarily be restricted in the range $0<c<1$ and can take
values outside this range.

\section{Stability analysis}
To perform a stability analysis of our dynamical system, we
linearize the system of equations (18) and (19) about the critical
points to get
\begin{eqnarray}
\frac{d\delta u}{dx}&=&3v_c\delta u+3(-1+u_c)\delta v,\\
\frac{d\delta v}{dx}&=&-\frac{3}{u_c^2}[\alpha v_c(c+v_c) +
Au_c^2(1+\alpha)] \delta u\\&\;& +
\frac{3}{u_c}[(c+2v_c)\alpha+u_c(1+2v_c+\alpha-A(1+\alpha))]\delta
v.\nonumber,
\end{eqnarray}
The eigenvalues of the above dynamical system (26) and (27)
corresponding to the three critical points Eqs. (20 - 25) are
\begin{eqnarray}
\lambda_{1}&=&\frac{3}{2(c-1)}(2+A-2c-Ac+\alpha+A\alpha-Ac\alpha\\
&\;&+\sqrt{4(-1+A(c-1))(c-1)^2(1+\alpha)+(2-2c+\alpha-A(c-1)(1+\alpha)^2)}
)\nonumber,
\end{eqnarray}
\begin{eqnarray}
\mu_{1}&=&\frac{-3}{2(c-1)}(-2-A+2c+Ac-\alpha-A\alpha+Ac\alpha\\
&\;&+\sqrt{4(-1+A(c-1))(c-1)^2(1+\alpha)+(2-2c+\alpha-A(c-1)(1+\alpha)^2)})\nonumber,
\end{eqnarray}
\begin{eqnarray}
\lambda_{2}&=&\frac{-3}{4}(-2-2\alpha-A(1+\alpha)-\sqrt{A}\sqrt{A+4c}(3+\alpha)+
\sqrt{2}[(2A^{5/2}\alpha+2(1+\alpha)^2\nonumber
\\&\;&+2\sqrt{A}\sqrt{A+4c}(1+\alpha)^2+2A(1+\alpha)(-1+c-2\sqrt{A+4c}+\alpha+c\alpha)\\&\;&+A^2(1-2\sqrt{A+4c}\alpha+\alpha^2)
+A^{3/2}(4-\sqrt{A+4c}+4\alpha-8c\alpha+\sqrt{A+4c}\alpha^2))]^{1/2})\nonumber,
\end{eqnarray}
\begin{eqnarray}
\mu_{2}&=&\frac{3}{4}(2+2\alpha+A(1+\alpha)+\sqrt{A}\sqrt{A+4c}(3+\alpha)+\sqrt{2}[(2A^{5/2}\alpha+2(1+\alpha)^2\nonumber\\&\;&
+2\sqrt{A}\sqrt{A+4c}(1+\alpha)^2+2A(1+\alpha)(-1+c-2\sqrt{A+4c}+\alpha+c\alpha)\\&\;&+A^2(1-2\sqrt{A+4c}\alpha+\alpha^2)
+A^{3/2}(4-\sqrt{A+4c}+4\alpha-8c\alpha+\sqrt{A+4c}\alpha^2))]^{1/2})\nonumber,
\end{eqnarray}
\begin{eqnarray}
\lambda_{3}&=&\frac{-3}{4}(-2-2\alpha-A(1+\alpha)+\sqrt{A}\sqrt{A+4c}(3+\alpha)+\sqrt{2}[(2A^{5/2}\alpha+2(1+\alpha)^2\nonumber\\&\;&
-2\sqrt{A}\sqrt{A+4c}(1+\alpha)^2+2A(1+\alpha)(-1+c-2\sqrt{A+4c}+\alpha+c\alpha)\\&\;&+A^2(1+2\sqrt{A+4c}\alpha+\alpha^2)
+A^{3/2}(4+\sqrt{A+4c}+4\alpha+8c\alpha-\sqrt{A+4c}\alpha^2))]^{1/2})\nonumber,
\end{eqnarray}
\begin{eqnarray}
\mu_{3}&=&\frac{3}{4}(2+2\alpha+A(1+\alpha)-\sqrt{A}\sqrt{A+4c}(3+\alpha)+\sqrt{2}[(2A^{5/2}\alpha+2(1+\alpha)^2\nonumber\\&\;&
-2\sqrt{A}\sqrt{A+4c}(1+\alpha)^2+2A(1+\alpha)(-1+c-2\sqrt{A+4c}+\alpha+c\alpha)\\&\;&+A^2(1+2\sqrt{A+4c}\alpha+\alpha^2)
+A^{3/2}(4+\sqrt{A+4c}+4\alpha+8c\alpha-\sqrt{A+4c}\alpha^2))]^{1/2})\nonumber.
\end{eqnarray}
It is easy to show that the real parts of the eigenvalues
($\lambda_1$, $\mu_1$) are negative, while for ($\lambda_2$,
$\mu_2$) and ($\lambda_3$, $\mu_3$), the real parts are all
positive. Hence a stable stationary attractor solution is possible
through the first critical point only. It can alternatively be
proved using the deceleration parameter $q$. Note that the
acceleration in the late evolution of the universe arises when
\begin{equation}
q=-\frac{\ddot{a}}{aH^2}\leq-1,
\end{equation}
which holds for the first critical point $(u_{1c},v_{1c})$ only,
since $q_1=-1$. While for $(u_{2c},v_{2c})$ and $(u_{3c},v_{3c})$,
we require
\begin{equation}
q_{2,3}=\frac{1}{2}\left[
1+\frac{3}{2}(A\mp\sqrt{A}\sqrt{A+4c})\right]<-1.
\end{equation}
As $A+4c\geq0$, the inequalities in Eqs. (35) do not hold and hence
the accelerated-expansion solution is not obtained from the second
and third critical points. Hence the valid attractor solution is
obtained from the first critical point only. We shall, henceforth,
deal with the first critical point only.

As shown in figure 1, the first critical point $(u_{1c},v_{1c})$ is
the stationary attractor solution for the interacting modified
Chaplygin gas with the coupling constant fixed at $c=0.5$. The
parameter $A$ can assume values in the range $-0.35\leq A\leq 0.025$
\cite{wu2}; while we choose $A=0.025$ for our numerical work. Also
note that if the parameter $\alpha<0$, then it yields a polytropic
equation of state for dark energy but for the MCG, we take
$\alpha=0.004$. It is evident that all the solutions of the
dynamical system with four different initial conditions converge to
the same final state. As $q_1=-1$, the first critical point gives
rise to an accelerated-expansion solution of the universe which is
consistent with the observations.

Moreover, the attractor solution corresponding to $(u_{1c},v_{1c})$
is also possible if $c$ takes values outside the usual considered
range of $0\leq c \leq1.$ In figures 2 and 3, the parameter $c$ is
given values $1.7$ and $-1.5$, respectively, with the same initial
conditions. Curiously, all the four solutions converge to the same
single final state. It draws to the fact, that at least
theoretically, the coupling constant $c$ can take values outside the
interval [0,1]. This result is further deduced in the next section
using a different formalism.

\section{Constraints on coupling constant}

We can determine the constraints on the coupling constant $c$ by
using the first critical point of our dynamical system. For this
purpose, we shall adopt the formalism of Guo and Zhang \cite{guo}.
We define new parameters corresponding to MCG and dark matter by
\begin{equation}
\gamma_{mcg}\equiv 1+\omega=\frac{\rho_{mcg}+p_{mcg}}{\rho_{mcg}},
\end{equation}
and
\begin{equation}
\gamma_{dm}\equiv\frac{\rho_{dm}+p_{dm}}{\rho_{dm}}.
\end{equation}
Note that $\gamma_{dm}=1$ since $p_{dm}=0$. Moreover, the parameter
$\gamma_{mcg}$ will be determined corresponding to the first
critical point. To find how the density ratio $R$ evolves with time,
we differentiate it with respect to $t$ to get
\begin{equation}
\dot{R}=\frac{dR}{dt}=\frac{\rho_{dm}}{\rho_{mcg}}\left[\frac{\dot{\rho}_{dm}}{\rho_{dm}}-\frac{\dot{\rho}_{mcg}}{\rho_{mcg}}\right].
\end{equation}
Using Eqs. (10) and (11), Eq. (38) becomes
\begin{equation}
\dot{R}=R\left[\frac{Q}{\rho_{dm}}+\frac{Q}{\rho_{mcg}}+
3H(\gamma_{mcg}-1)\right].
\end{equation}
Using Eq. (12) in (39), we get after simplification
\begin{equation}
\dot{R}=3H[c(1+R)^2 +R(\gamma_{mcg}-1)].
\end{equation}
In order to get stationary solutions, we solve for $\dot{R}=0$ to
get
\begin{equation}
R^\pm_s=\frac{1-\gamma_{mcg}}{2c}-1\pm\sqrt{\left(\frac{1-\gamma_{mcg}}{2c}-1\right)^2-1}.
\end{equation}
Now, to get real valued solutions, we require
$(\frac{1-\gamma_{mcg}}{2c}-1)^2-1\geq0$, which yields
\begin{equation}
\left(\frac{1-\gamma_{mcg}}{2c}-2\right)\left(\frac{1-\gamma_{mcg}}{2c}\right)\geq0.
\end{equation}
The above inequality holds if the quantities in the brackets are
either both positive or both negative. We shall take $c$ to be a
free parameter which can take values other than zero.

Case (1)

Assume both quantities in the brackets in (42) to be positive, i.e.
\begin{equation}
\frac{1-\gamma_{mcg}}{2c}-2\geq0, \ \
\frac{1-\gamma_{mcg}}{2c}\geq0.
\end{equation}

Case (1a)

 Now take $c>0$; thus Eq. (43) gives
\begin{equation}
\gamma_{mcg}\leq1-4c, \ \ \gamma_{mcg}\leq1,
\end{equation}
which yields
\begin{equation}
\gamma_{mcg}\leq1-4c.
\end{equation}

Case (1b)

 If $c<0$, then Eq. (43) yields
\begin{equation}
\gamma_{mcg}\geq1-4c, \ \ \gamma_{mcg}\geq1,
\end{equation}
which implies
\begin{equation}
\gamma_{mcg}\geq1-4c.
\end{equation}

Case (2)

 Now take both quantities in the brackets in Eq. (42) to be
negative, i.e.
\begin{equation}
\frac{1-\gamma_{mcg}}{2c}-2\leq0, \ \
\frac{1-\gamma_{mcg}}{2c}\leq0.
\end{equation}

Case (2a)

 Take $c>0$; after solving Eq. (48), which gives
\begin{equation}
\gamma_{mcg}\geq1-4c, \ \ \gamma_{mcg}\geq1,
\end{equation}
which yield
\begin{equation}
\gamma_{mcg}\geq1.
\end{equation}

Case (2b)

 If $c<0$, then Eq. (48) yields
\begin{equation}
\gamma_{mcg}\leq1-4c, \ \ \gamma_{mcg}\leq1,
\end{equation}
which results in
\begin{equation}
\gamma_{mcg}\leq1.
\end{equation}
Now we shall use the definition
$\gamma_{mcg}=1+\omega_1=1+v_{1c}/u_{1c}$ in each of the above four
cases.

Case (1a)

 Using $\omega_1=\frac{v_1}{u_1}=\frac{-1}{1-c}$ in Eq.
(45), we have $(1-2c)^2\geq0$ ,which is satisfied for all values of
$c$. Notice that from Eq. (44), we have an additional constraint
$c<1$; therefore, $0<c<1$, which is the range usually considered for
$c$ in the literature.

Case (1b)

 Using $\omega_1$ in Eq. (47) we get $(1-2c)^2\leq0$, which
is satisfied only for $c=1/2$. Since $c<0$, we do not have an
acceptable solution.

Case (2a)

 Here for $\omega_1$, Eq. (50) implies
$\frac{-1}{1-c}\geq0$, which holds for all $c>1$. Apparently it
implies that the coupling constant between MCG and dark matter can
take arbitrary value; thus, the mutual interaction can be more
dynamic. It yields arbitrary freedom for the conversion of MCG into
dark matter.

Case (2b)

 For $\omega_1$, Eq. (52) implies $\frac{-1}{1-c}\leq0$,
which is viable if $c<0$. Thus, the coupling constant can take
arbitrary negative values. This case apparently supports the
conversion of dark matter into MCG with arbitrary coupling. Note
that $\omega_1$ represents the EoS of phantom energy
($\omega_1<-1$).

\section{Conclusion}
We have investigated the possible interaction between dark matter
and the Chaplygin gas and we deduced that the coupling constant
involved can take values outside the range usually considered,
$0<c<1$. This range arises as a special case in the Case (1a). Our
analysis suggests that $c$ can take arbitrary positive or negative
values. If $c>1$, as in Case (2a), then it supports the conversion
of MCG into dark matter. Conversely, if $c<0$ as in Case (2b), it
allows for the conversion of dark matter into MCG. It also supports
the existence of phantom energy through the final case. Moreover,
the present work may serve as the generalization of the earlier work
by Zhang and Zhu \cite{zhang} for the interacting Chaplygin gas and
by Wu and Yu \cite{wu} for the interacting generalized Chaplygin
gas.

In a recent investigation, Feng et al \cite{feng2} have presented
observational constraints on the coupling parameter and have deduced
that small positive values for $c$ are most probable. This
conclusion is drawn in order to alleviate the cosmic-coincidence
problem. Also the negative values of $c$ are excluded to avoid the
violation of the second law of thermodynamics \cite{pavon}. We have
deduced from our analysis that $c$ has no such theoretical
constraints, and the usual choice [0,1] is not a true range for the
coupling parameter.

\section*{Acknowledgements} One of us (MJ) would like to thank A. Qadir, S. Odintsov, N. Riazi and H. Zhang for
useful discussions during this work. We would also like to thank the
anonymous referee for giving useful comments, which helped in
improving the paper. 
\begin{figure}
\includegraphics{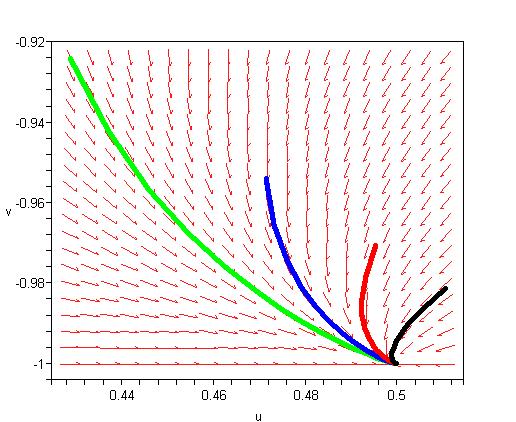}\\
\caption{The phase diagram of the interacting modified Chaplygin gas
model with $c=0.5$. The model parameters are fixed as $A=0.025$ and
$\alpha=0.004$. The curved lines from left to right correspond to
the initial conditions $u(-2)=1.2, v(-2)=-0.2$ (green); $u(-2)=1.3,
v(-2)=-0.3$ (blue); $u(-2)=1.4, v(-2)=-0.4$ (red); $u(-2)=1.5,
v(-2)=-0.5$ (black).}
\end{figure}
\newpage
\begin{figure}
\includegraphics{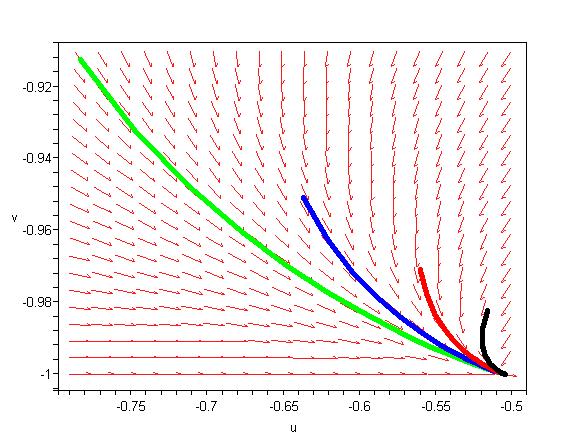}\\
\caption{ The phase diagram of the interacting modified Chaplygin
gas model with $c=1.7$. The model parameters are fixed as in Fig. 1.
The curved lines from left to right correspond to the initial
conditions as given in Fig. 1.}
\end{figure}
\newpage
\begin{figure}
\includegraphics{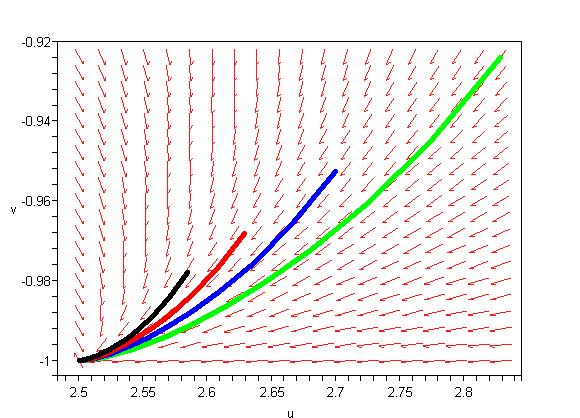}\\
\caption{ The phase diagram of the interacting modified Chaplygin
gas model with $c=-1.5$. The model parameters are fixed as in Fig.
1. The curved lines from left to right correspond to the initial
conditions as given in Fig. 1.}
\end{figure}

\end{document}